\documentstyle[12pt]{article}

\catcode`\@=12

\relax
\def\figcap{\section*{Figure Captions\markboth
	{FIGURECAPTIONS}{FIGURECAPTIONS}}\list
	{Fig. \arabic{enumi}:\hfill}{\settowidth\labelwidth{Fig. 999:}
	\leftmargin\labelwidth
	\advance\leftmargin\labelsep\usecounter{enumi}}}
 \relax
\def\reflist{\section*{References\markboth
	{REFLIST}{REFLIST}}\list
	{[\arabic{enumi}]\hfill}{\settowidth\labelwidth{[999]}
	\leftmargin\labelwidth
	\advance\leftmargin\labelsep\usecounter{enumi}}}
 \relax

\begin{document}

\newcommand{\npb}[3]{Nucl. Phys.{\bf B#1} (#2) #3}
\newcommand{\plb}[3]{Phys. Lett. {\bf B#1} (#2) #3}
\newcommand{\prd}[3]{Phys. Rev. {\bf D#1} (#2) #3}
\newcommand{\prl}[3]{Phys. Rev. Lett. {\bf #1} (#2) #3}
\newcommand{\prp}[3]{Phys. Rep. {\bf #1} (#2) #3}

\begin{flushright}
 CCAST 94-02

hep-ph/9408283
\end{flushright}

\begin{center}

{ \bf \Large QCD Corrections to Charmless $b \to s$ Processes\footnote{
This work is partly supported by the National
Natural Science Foundation of China and the Doctoral Program Foundation
of Institution of  Higher Education.}
}

{ \bf \large Chong-Shou Gao$^{a,b,c}$, Cai-Dian
L\"{u}$^{a,b,c}\footnote{E-mail: lucd@itp.ac.cn.}$
	 and Zhao-Ming Qiu$^d$ \\
a CCAST(World Laboratory), P.O.Box 8730, Beijing 100080, China\\
b Institute of Theoretical Physics, Academia Sinica, P.O. Box 2735,\\
Beijing 100080, China\\
c Physics Department, Peking University,
Beijing 100871, China\\
d General Office, Chinese Academy of Sciences}

\begin{abstract}
QCD corrections to the penguin induced charmless inclusive
$B$ decays are performed to an improved leading-log approximation.
For process $b \to s q\overline{q} $ (q=uds) and $ b \overline{q}
\to s \overline{q} $, four previous missed 4-quark operators are
included to make a complete leading log QCD corrected result.
Furthermore, part of the next-leading log effect was also given for
the above processes and process $b \to s g$.
In comparison to previous QCD corrected calculations by
Grigjanis et al., the inclusive decay rates for process $b \to s
q\overline{q} $ (q=ud), $b\to s s\overline{s}$ and $ b \overline{q}
\to s \overline{q} $ are suppressed 10\%, 14\% and 18\% respectively.
The branching ratio of process $b \to s g$ is enhanced by 2-3 orders,
the result for this process is in agreement with the result of Ciuchini et al.
\end{abstract}

{\it Submitted to Z. Physik C}

\end{center}

\vfill

\newpage

		\section{Introduction}
\label{intro}

Recently the CLEO collaboration has observed
 the charmless decays of $B$, $B^0 \to \pi^+\pi^-$,
$B^0 \to K^+\pi^-$ and $B^0 \to K^+ K^-$\cite{cleo3}.
The upper limits on the branching fractions are
$B_{\pi\pi} < 2.9 \times
10^{-5}$, $B_{K\pi} < 2.6 \times 10^{-5}$, $B_{KK} < 0.7 \times
10^{-5}$. The sum of $B_{\pi\pi}$ and $B_{K \pi}$ exceeds zero
with a significance more than four standard deviations.
One can expect,  individuals of these exclusive channels
and also the inclusive processes will be found at CESR or future
B factories.

The B meson system is expected to give rise to a very rich phenomenology,
providing a wealth of information that will allow us to constrain the currently
very successful standard model (SM), as well as many extensions that go
beyond it. It also provides many channels for the search of CP violation.

The charmless $b$ quark decay has already been calculated
 in several  papers\cite{Hou,Gri,Des,Cha,Hou2}.
They are induced by both $b\to u$ transition and penguin diagrams.
The $b\to u$ transition diagrams are easy to calculate, but
penguin diagrams are more complicated. W.S. Hou showed the
results of penguin diagrams without QCD corrections in ref.\cite{Hou}.
Afterwards, results including some
short distance QCD corrections were given by R. Grigjanis et
al.\cite{Gri}. For different processes, the authors claimed that
the branching ratios are either QCD-enhanced or suppressed by a notable
factor. In other words, the strong interaction plays an important
role in these decays. But this is not the end of the story, there
are still some unsolved questions.
The anomalous dimensions used in ref.\cite{Gri} were changed later
by the authors themselves in ref.\cite{Gri2}'s erratum. This change caused
large effects in process $b\to sg$ which can be seen in the
present paper. See also a recent work by Ciuchini et al.\cite{ciu2}.
 Furthermore, only two
four-quark operators were considered there, while the other four were
neglected.
Unfortunately, the missing four operators are responsible for contributing
to the matrix elements of
processes $b \to s q\overline{q} $ (q=uds) and $ b \overline{q} \to s
\overline{q} $. Although their coefficients of matching at $M_W$ can
be transferred to the operator $O_7$ of ref.\cite{Gri}\footnote{
Note this operator $O_7$ is not the one that we defined in this paper},
when running, these six four-quark
operators will be mixed together, so the missing four will reappear with
nonzero coefficients. This is similar to the $b\to s\gamma$
case.\footnote{See discussions in ref.\cite{Gri2,Mis2}.}

Since charmless decays of $B$ meson caused a great interest
in studying CP violation, and the present experimental limit
falls in with the theoretical predictions of the standard model, a more
accurate calculation is needed to reduce theoretical uncertainties. The
 aim of the present paper is to present an improved
leading log QCD corrected result.

\section{Matching at $M_W$}
\label{mu=w}

 The basic effective field theory idea is by now quite well
established\cite{Witt,Georgi}.
In our particular case of Minimal Standard Model, we integrate
out the top quark and the weak W bosons at $\mu=M_W$ scale,
generating an effective five-quark theory. By using the
renormalization group equation, we run the effective field theory
down to b-quark scale to calculate QCD corrections to charmless $b$ decay.

After applying the full QCD equations of motion\cite{EOM}, a complete
set of operators relevant for charmless $B$ decays can be chosen to be:
\begin{equation}
\begin{array}{ll}
O_1=(\overline{c}_{L\beta} \gamma^{\mu} b_{L\alpha})
	    (\overline{s}_{L\alpha} \gamma_{\mu} c_{L\beta}),
& O_2=(\overline{c}_{L\alpha} \gamma^{\mu} b_{L\alpha})
	    (\overline{s}_{L\beta} \gamma_{\mu} c_{L\beta}),\\
O_3=(\overline{s}_{L\alpha} \gamma^{\mu} b_{L\alpha})
	\displaystyle{\sum _q} (\overline{q}_{L\beta} \gamma_{\mu}
	q_{L\beta}),
& O_4=(\overline{s}_{L\alpha} \gamma^{\mu} b_{L\beta})
	\displaystyle{\sum _q} (\overline{q}_{L\beta} \gamma_{\mu}
	q_{L\alpha}),\\
O_5=(\overline{s}_{L\alpha} \gamma^{\mu} b_{L\alpha})
	 \displaystyle{\sum _q} (\overline{q}_{R\beta}
	\gamma_{\mu} q_{R\beta}),
& O_6=(\overline{s}_{L\alpha} \gamma^{\mu} b_{L\beta})
	\displaystyle{\sum _q} (\overline{q}_{R\beta}
	\gamma_{\mu} q_{R\alpha}),
\end{array}
\end{equation}
$$O_7=(g_3/16\pi^2) m_b \overline{s}_{L} \sigma^{\mu\nu}
	    T^a b_{R} G_{\mu\nu}^a.$$
The covariant derivative is defined as
$$D_{\mu}=\partial_{\mu}-i\mu^{\epsilon/2}g_3 X^a G_{\mu}^{a},$$
with $g_3$ denoting the QCD coupling constant.

Then we can write down our effective Hamiltonian as
\begin{equation}
{\cal H}_{eff}=2 \sqrt{2} G_F V_{tb}V_{ts}^*\displaystyle \sum _i
C_i(\mu)O_i(\mu). \label{eff}
\end{equation}
One can find the coefficients of operators at $\mu=M_W$ scale
by integrating out the weak gauge bosons and would-be
Goldstone bosons at this scale. Since in standard model, there is no
tree level flavor changing neutral current, all charmless $b\to s$ processes
are through loop diagrams. They are QCD induced processes. Even if
no radiative QCD corrections are performed, they are already order of
O($\alpha_s$)\cite{Hou}. The decay amplitudes of four-quark processes
are proportional to
$[f_1 \alpha_s \log(M_W/m_b)^2 + f_2\alpha_s]$. The leading logarithmic QCD
corrections to them are of O($[\alpha_s \log (M_W/m_b)^2 ]^n$) with
$n\ge 2$. They are summed by the one-loop renormalization group equation
in the leading log approximation. So this approximation will include the
leading log terms of all loops but miss the $f_2\alpha_s $ terms of
one-loop diagrams. Although this is usually considered as part of
the next-to-leading
 log effect, it is actually part of the one-loop result. We will use
the complete one-loop result (including $f_2 \alpha_s$) for four-quark
operators in matching at $M_W$, so that we can get a result from both
the complete one-loop contribution and summation of leading
log contribution. This is an improvement
in the perturbative calculation at least for the QCD-induced four-quark
processes. The matching diagrams for four-quark operators  are displayed
in Fig.\ref{fmat3} (There should also be three additional symmetric
four quark diagrams at each side of the first equation in Fig.1.). Matching
diagrams for operator $O_7$ are very similar to $b\to s \gamma$
calculations\cite{gri3}.
Neglecting small terms proportional to $m_c^2$ or $m_u^2$
in these matching conditions, and using $V_{cb}V_{cs}^* = - V_{tb}V_{ts}^*$,
one finds the following coefficients of the operators:\cite{Mis2,Bur}
\begin{eqnarray}
C_1(M_W) &=& \frac{ 11 \alpha_s(M_W) }{8\pi},
	~~~~C_2(M_W) ~=~ 1-\frac{ 11 \alpha_s(M_W) }{24\pi},  \nonumber\\
C_3(M_W)  &=& C_5(M_W) ~=~ \frac{ \alpha_s(M_W) }
	{24\pi} F_1(\delta)\nonumber\\
C_4(M_W)  &=& C_6(M_W) ~=~ -\frac{ \alpha_s(M_W) }
	{8\pi} F_1(\delta)\nonumber\\
C_{ 7}(M_W) &=& \frac{ -\frac{1}{8} +\frac{5}{8}\delta +\frac{1}{4}\delta^2 }
	{ (1-\delta)^3 } +\frac{ \frac{3}{4}\delta^2 }{ (1-\delta)^4 }
	\log\delta
\end{eqnarray}
where
\begin{equation}
F_1(\delta) = \frac{2}{3} +\frac{ -\frac{1}{12} -\frac{11}{12}\delta
	+\frac{3}{2}\delta^2 }{ (1-\delta)^3 }
	+\frac{ -\frac{3}{2}\delta^2 +\frac{8}{3}\delta^3
	-\frac{2}{3}\delta^4 }{ (1-\delta)^4 } \log \delta
\end{equation}
with $\delta =M_W^2 / m_t^2$.

\section{Renormalization group running from $M_W$ to $m_b$}
\label{mwb}

We then use the renormalization group equation satisfied by
the coefficient functions $C_i(\mu)$,
to continue running
the coefficients of operators from $\mu=M_W$ to $\mu=m_b$.
\begin{equation}
\mu \frac{d}{d\mu} C_i(\mu)=\displaystyle\sum_{j}(\gamma^{\tau})_{
ij}C_j(\mu),\label{ren}
\end{equation}
Where $\gamma_{ij}^,s$ are anomalous dimensions of operators. These anomalous
dimensions have been calculated by many authors for the process $b\to s
\gamma$.
Only recently it is completely solved by Ciuchini et al.\cite{Ciu},
and their calculation is confirmed by Cella et al.\cite{Cel}.
\begin{equation}
\begin{array}{ccc}
\gamma = &
	\left( \begin{array}{ccccccc}
	-1 &  3 &     0   &   0  &     0  &    0    &   3/2\\
	3 &  -1 &   -1/9  &  1/3 &   -1/9 &   1/3   &   38/27 \\
	0 &   0 &   -11/9 & 11/3 &   -2/9 &   2/3   &   557/54 \\
	0 &   0 &    22/9 &  2/3 &   -5/9 &   5/3   &   271/27 \\
	0 &   0 &     0   &    0 &     1  &   -3    &   -37/6 \\
	0 &   0 &   -5/9  &  5/3 &   -5/9 &   -19/3 &   -673/54 \\
	0 &   0 &     0   &    0 &     0  &    0    &   14/3
	\end{array}\right)
      & \displaystyle{ \frac{g_3^2}{8 \pi ^2} }\label{anom}
\end{array}
\end{equation}

\begin{table}
\begin{center}
\caption{Numerical results for coefficients of operators
$C_i(m_b)$ with $\alpha_s(m_Z)=0.117$.}
\begin{tabular}{|c|c|c|c|c|c|}
\hline
$m_{top}$(GeV) & {$C_3(m_b)$}
      & {$C_4(m_b)$} & {$C_5(m_b)$} & {$C_6(m_b)$ } & $C_7(m_b)$ \\
\hline
100    &  0.012   & -0.027  & 0.008  & -0.034  & -0.158\\
\hline
140    &  0.012   & -0.028  & 0.008  & -0.035  & -0.170\\
\hline
180    &  0.013   & -0.028  & 0.008  & -0.035  & -0.177\\
\hline
220    &  0.013   & -0.028  & 0.008  & -0.036  & -0.182\\
\hline
260    &  0.013   & -0.028  & 0.008  & -0.036  & -0.185\\
\hline
300    &  0.013   & -0.029  & 0.008  & -0.036  & -0.187\\
\hline
\end{tabular}\label{Cmb}
\end{center}
\end{table}

In the leading order of $g_3$, the solution to eqn.(\ref{ren})
in  matrix notation is given by
\begin{equation}
C(\mu_2)=\left[\exp\int_{g_3(\mu_1)}^{g_3(\mu_2)}dg\frac
{\gamma^T(g)}{\beta(g)}\right] C(\mu_1).\label{solu}
\end{equation}
 After insertion of anomalous dimension matrix(\ref{anom}), we get
the coefficients of operators at $\mu=m_b$ scale.
Here we made use of $M_W=80.22$GeV, $m_b=4.9$GeV,
$\alpha_s(m_Z)=0.117$\cite{data}.
Table~\ref{Cmb} shows the numerical
values of coefficients of operators $O_i$, with different
input of top quark mass.  In this table, coefficients of four-quark
 operators $O_{3, 4,5,6}$ have very little dependence on the top quark mass,
while $C_7(m_b)$ is getting bigger when top mass increases.
$C_2(m_b)=1.077$. Its coefficient does not vary with top mass at all,
because operator $O_2$ is generated from $W$ exchange diagram and does not
mix with top quark loop induced operators,

\section{The charmless b decay rate}
\label{rrate}

	There are four types of nonleptonic charmless decays considered
here. Their inclusive decay widths are
given by the sum of operators $O_i$ in our effective field theory.

(1) Operator $O_5$ and $O_7$ in our operator
basis contribute to  process $ b \to s g$\cite{Ciu}.
The decay width is,
\begin{equation}
\Gamma_{b\to s g} = \frac{8\alpha _s}{\pi}
\left| C_7^{eff} (m_b) \right|^2 \Gamma _0,
\end{equation}
where $$\Gamma _0 = \frac{G_F^2 m_b^5 }{ 192\pi^3 } |V_{ts}^* V_{tb} |^2 ,$$
$$ C_7^{eff}(m_b) = C_7(m_b) +C_5(m_b).$$

(2) For process
$ b \to s q \overline{q} $($q$ denotes $u$ or $d$ quark), the Feynman
diagrams are displayed in Fig.\ref{rate}. Operators $O_{2,3,4,5,6,7}$
contribute to this process in the effective  field theory.
The decay rate is,
$$\Gamma_{b\to s q \overline{q} } = \left[ 3 \displaystyle{ \sum _{i=3}
^{6} } \left| C_i^{eff}(m_b) \right|^{2}
+2 \left\{ C_3^{eff}(m_b)  C_4^{eff}(m_b)  +C_5^{eff}(m_b)
  C_6^{eff}(m_b)  \right\} \right.$$
\begin{equation}
\left. +\frac{8\alpha_s }{3\pi}
\left\{ C_4^{eff}(m_b)  +C_6^{eff}(m_b)  \right\}C_7(m_b) \right] \Gamma_0,
\end{equation}
where
$$C_i^{eff} (m_b) =C_i(m_b)
- \frac{\alpha_s}{36\pi} \log\frac{m_c^2}{m_b^2} ~C_2(m_b) ,~~~~ i=3,5, $$
$$C_i^{eff} (m_b) =C_i(m_b)
+\frac{\alpha_s}{12\pi} \log\frac{m_c^2}{m_b^2} ~C_2(m_b) , ~~~~i=4,6 ,$$
with $m_c=1.5GeV$. Here we included all terms proportional to $\alpha_s$,
to make a complete one-loop O($\alpha_s$) result.

(3) For process
$ b \to s s \overline{s} $, there are additional diagrams concerning
momentum exchange of two s quarks.
The decay width is,

\begin{equation}
\Gamma_{b\to s s \overline{s} } = \left[ 4 \left|C_3^{eff} (m_b)
+C_4^{eff} (m_b)\right|^{2}
+3\left|C_5^{eff}(m_b) \right|^{2} +3\left| C_6^{eff}(m_b)
 \right|^{2}  \right.
\end{equation}
$$\left.+2 C_5^{eff}(m_b) C_6^{eff}(m_b)
 +\frac{8\alpha_s }{3\pi}
\left\{ C_3^{eff}(m_b)  +C_4^{eff}(m_b)  +C_6^{eff}(m_b)
 \right\}C_7(m_b)  \right] \Gamma_0.$$

(4) The Feynman diagrams contributing to the nonspectator process
$b \overline{q} \to s \overline{q} $ are also seen in Fig.\ref{rate}.
The decay width is,
\begin{equation}
\Gamma_{b \overline{q} \to s \overline{q} } =
32 \pi^2  \left|C_5^{eff} (m_b) +3 C_6^{eff} (m_b)\right|^{2}
\left( \frac{ f_{B} }{ m_b} \right)^2 \Gamma_0.
\end{equation}
Where the B decay constant is $f_B=200MeV$.

	In order to find the branching ratios of these kinds of
charmless B decays, the semileptonic decay of B is used.
\begin{equation}
BR(\overline{B} \rightarrow X_s ~no~charm )=
\frac{ \Gamma( b \rightarrow s ~no ~charm) }{  \Gamma(b
\rightarrow c e \overline{\nu} ) }  BR( \overline{B} \rightarrow X_c
 e \overline{\nu}),
\end{equation}
where
\begin{equation}
\Gamma(b \rightarrow c e \overline{\nu})  \simeq
g (m_c/m_b) \left(1-\frac{2 \alpha_{s}(m_b)}{3 \pi} f(m_c/m_b) \right)
\Gamma_0,
\end{equation}
with $g(m_c/m_b)\simeq 0.447$ and $f(m_c/m_b)\simeq 2.4$ correspond
to the phase space
factor and the one-loop QCD correction to the semileptonic decay, respectively
\cite{Cabi}. Here we use experimental result $Br(\overline{B} \to
X_c e\overline{\nu} ) =11\% $\cite{data}.

The branching ratios for different processes are given in Fig.~\ref{bsg}.
In comparison to the previous QCD-corrected results of ref.\cite{Gri},
the decay rate of process $b \to sg$
is strongly enhanced by QCD corrections rather than severely suppressed.
The differences are of 2-3 orders. The reason for such a large
difference, is obvious, for the authors of ref.\cite{Gri} used the wrong
anomalous dimensions of ref.\cite{Gri2}, which has already been corrected
in its erratum. If right values are taken, the differences are not so large.
Our result for $b\to s g$ is consistent with the recent work by Ciuchini
et al.\cite{ciu2} in leading log approximation. The part of next-to-leading
log contributions we considered in this paper
is not essential to this process.
For process $b \to s q\overline{q}$,
$b \to s s\overline{s}$ and $b\overline{q} \to s \overline{q}$,
after we include all the dimension-6 four-quark operators, the
complete leading log QCD corrections make the branching ratios suppressed
by 15\%, 19\%, and 25\% respectively (comparing to the result of
 ref.\cite{Gri}), when
 $m_{top}=175GeV$. Furthermore, we also include O($\alpha_s$)
terms in the matching condition, which is the first term of next-to-leading
log effects. The branching ratios for these processes are slightly enhanced.
Comparing with ref.\cite{Gri},
the total effects are 10\%, 14\%, and 18\% suppression of the branching
ratios, respectively,

 The branching ratios obtained by W.S. Hou\cite{Hou}
without QCD running from $M_W$ to $m_b$ are also given at Fig.~\ref{bqsq}.
Comparison with this result shows that
 our decay rate of process $b \to sg$ is strongly
enhanced, e.g. a factor of 3.4 at $m_{top}=175GeV$, this is almost the same
 as $b\to s\gamma$ decay\cite{gri3,Ciu}.
While for process $b \to s q\overline{q}$,
$b \to s s\overline{s}$ and $b\overline{q} \to s \overline{q}$,
the branching ratios are only increased by 49\%, 34\% and 17\%,
respectively.

\section{Conclusion}
\label{conclusion}

In conclusion, we have given the full leading log QCD corrections to
penguin induced charmless $b\to s$ processes together with some next-leading
log corrections.
The branching ratios of processes $b \rightarrow s q\overline{q}$
(q=uds), and $b\overline{q} \to s \overline{q}$ are found to
change slowly with the
top quark mass. For process $b \to s g$, the decay rate is getting larger
when top mass increases. The  result also shows
that for $b\to sg$ process(gluon on shell), QCD corrections are
as important as in the $ b\to s \gamma$ case. The QCD corrected results
for four-quark processes are all enhanced, the amount of the increase vary
notably for different processes.

\bigskip
\bigskip
{\noindent \bf { Acknowledgment}}
\bigskip

One of the authors(C.D. L\"{u}) thanks Prof. Xiaoyuan Li
and Dr. Y.Q. Chen, Y. Liao, Q.H. Zhang for
helpful discussions.


\begin{figcap}

\item{Matching conditions at $\mu=M_W$ for
Green functions in the full standard model(left hand side)
and effective field theory below W scale(right hand side)
with the heavy dots denoting high dimension operators.}
\label{fmat3}

\item{Feynman diagrams contributing to $b \to s q\overline{q}$, $b
\overline{q} \to s \overline{q}$ in effective field theory. }
\label{rate}

\item{QCD corrected branching ratios
as function of top quark mass. From top to bottom are
lines for process $b \to s g$, $b \to s q\overline{q}$,
$b \to s s\overline{s}$ and $b \overline{q} \to s \overline{q}$.}
\label{bsg}

\item{Branching ratios without QCD radiative corrections,
obtained by W.S. Hou
as function of top quark mass. From top to bottom are
lines for process $b \to s q\overline{q}$,
$b \to s s\overline{s}$, $b \overline{q} \to s \overline{q}$
and $b \to s g$.}
\label{bqsq}
\end{figcap}

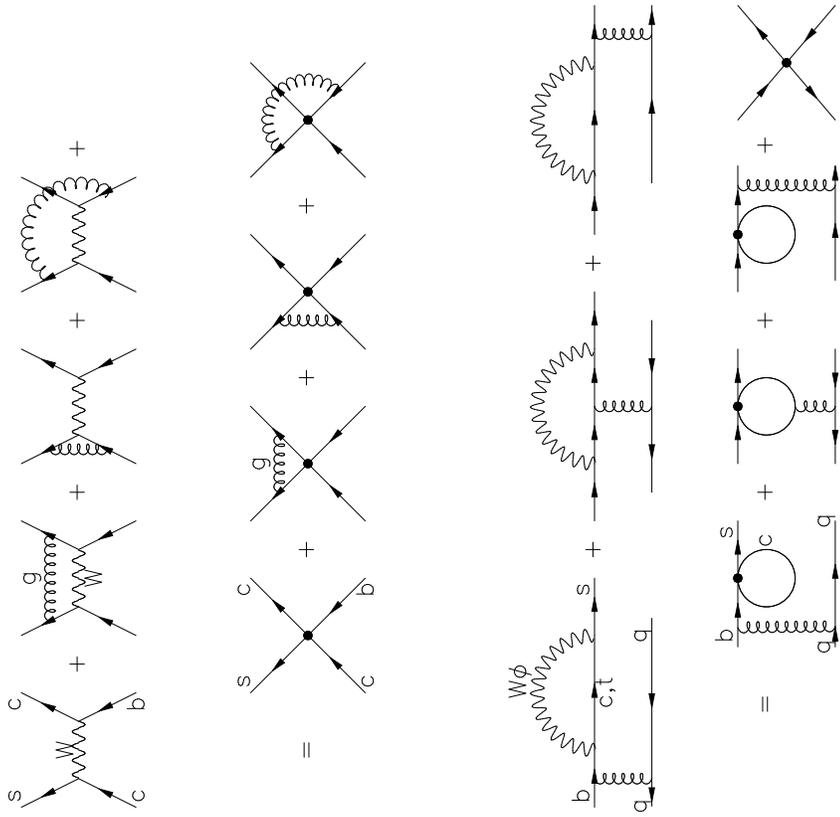
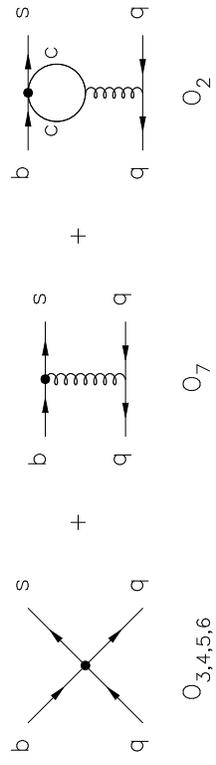
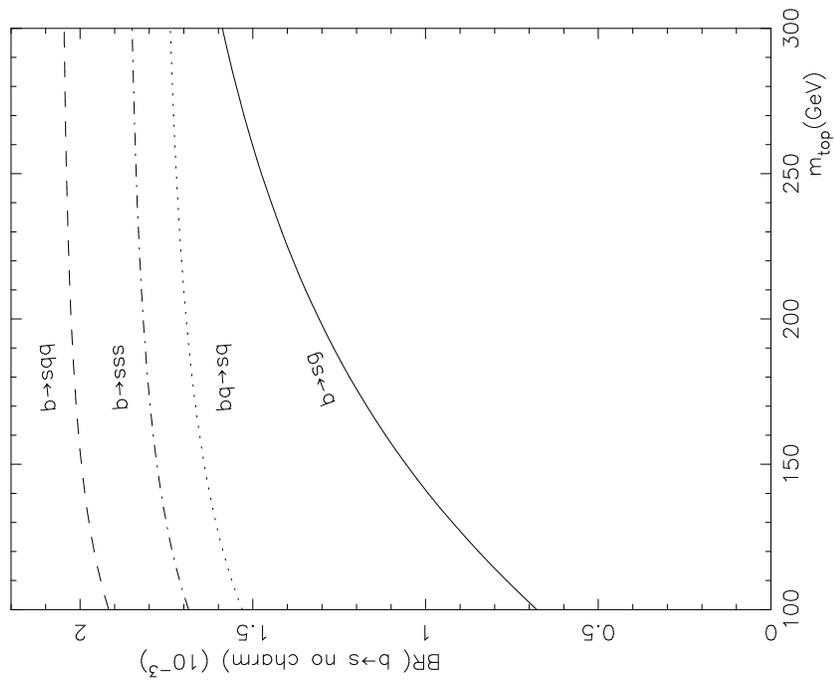
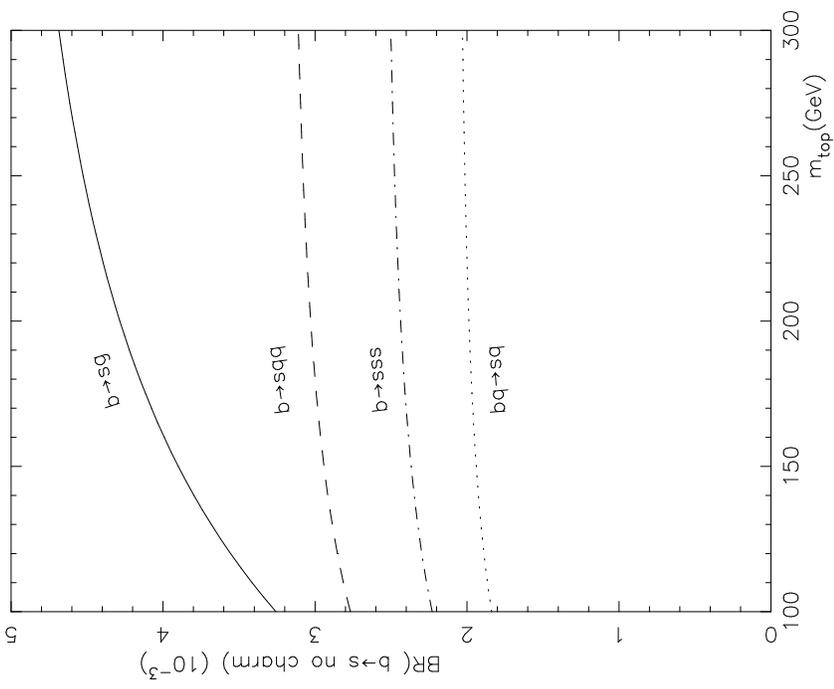
\end{document}